\documentclass[reprint, amsmath,amssymb, aps]{revtex4-1}

\usepackage{float}
\usepackage{graphicx}
\usepackage{dcolumn}
\usepackage{bm}
\usepackage{ulem}
\graphicspath{{fig/}}
\usepackage[colorlinks=true,linkcolor=blue,citecolor=blue, urlcolor=blue]{hyperref}

\begin{document}


\title{Photoproduction of doubly heavy baryons at future $e^+e^-$ colliders}

\author{Xi-Jie Zhan}
 	\email{zhanxj@cqu.edu.cn}

\author{Xing-Gang Wu}%
	\email{wuxg@cqu.edu.cn}

\author{Xu-Chang Zheng}
	\email{zhengxc@cqu.edu.cn}

\affiliation{Department of Physics, Chongqing Key Laboratory for Strongly Coupled Physics, Chongqing University, Chongqing 401331, People's Republic of China}

\date{\today}

\begin{abstract}

The photoprodution of doubly heavy baryon ($\Xi_{cc},\Xi_{bb},\Xi_{bc}$) is investigated in the context of future high-energy and high-luminosity $e^+e^-$ colliders.
The study incorporates two sources of initial photons, namely the LBS photon and the WWA photon. Alongside the direct photoproduction via the sub-process $\gamma+\gamma \rightarrow \Xi_{QQ^{'}} +\bar{Q}+\bar{Q^{'}}$ ($Q^{(')}=c,b$), the resolved photoproduction channels are specifically considered, encompassing the sub-processes $\gamma + g \rightarrow \Xi_{QQ^{'}} +\bar{Q}+\bar{Q^{'}}$, $g + g \rightarrow \Xi_{QQ^{'}} +\bar{Q}+\bar{Q^{'}}$, and $q + \bar{q} \rightarrow \Xi_{QQ^{'}} +\bar{Q}+\bar{Q^{'}}$ with $q=u,d,s$.
Within the framework of non-relativistic QCD, two $(cc(bb))$-diquark configurations, ${}_{\bar{\textbf{3}}}[{}^3S_1]$ and ${}_{\textbf{6}}[{}^1S_0]$, and four $(bc)$-diquark configurations, $(bc)_{\bar{\textbf{3}}}[{}^3S_1]$, $(bc)_{\textbf{6}}[{}^1S_0]$, $(bc)_{\textbf{6}}[{}^3S_1]$ and $(bc)_{\bar{\textbf{3}}}[{}^1S_0]$, are considered in the calculations.
Numerical results show that the single resolved photoproduction processes provide dominant contributions under certain collision configuration. At the future $e^+e^-$ colliders, the doubly heavy baryon generated via the photoproduction mechanism is promisingly observable and can be well studied.
\end{abstract}

\maketitle

\section{\label{sec:1}Introduction}
Baryon containing two heavy quarks, referred to as doubly heavy baryons, offers a simplified structure akin to heavy quarkonia, thus enabling rigorous theoretical analysis.
The first suspected observation of $\Xi_{c c}^+$ was reported by the SELEX Collaboration \cite{Mattson:2002vu,Ocherashvili:2004hi} in 2002 and 2005.
Lately in 2017, the LHCb Collaboration identified another doubly heavy baryon, $\Xi_{cc}^{++}$, through the decay mode $\Xi_{cc}^{++} \rightarrow \Lambda_{c}^{+} K^{-} \pi^{+} \pi^{+}$, with $\Lambda_{c}^{+} \rightarrow p K^{-} \pi^{+}$ \cite{Aaij:2017ueg}.
Further validation came from the LHCb Collaboration, confirming this baryon's existence via the decay channel $\Xi_{cc}^{++} \rightarrow \pi^+ \Xi_{c}^+$ \cite{Aaij:2018gfl, Aaij:2018wzf}. These observations render the doubly heavy baryon a valuable environment for investigating quantum chromodynamics (QCD).
Due to its nonrelativistic nature and the strong interaction confinement, the production of doubly heavy baryons involves nonperturbative effects that cannot be calculated using perturbative QCD. In the work by Ma et al. \cite{Ma:2003zk}, the nonrelativistic QCD (NRQCD) \cite{Bodwin:1994jh} factorization framework was employed to describe the production process. This framework divides the process into two stages: the perturbative creation of a heavy-quark pair in a certain quantum state, referred to as a diquark, followed by its nonperturbative transition into a baryon. By expanding in the small velocity ($v_Q$) of the heavy quark in the baryon's rest frame, two leading-order states of $(cc)$-diquarks were identified: ${}_{\bar{\textbf{3}}}[{}^3S_1]$ and ${}_{\textbf{6}}[{}^1S_0]$, each associated with a corresponding long-distance matrix element (LDME), namely $h_{\bar{\textbf{3}}}$ and $h_{\textbf{6}}$. 
${}_{\bar{\textbf{3}}}[{}^3S_1]$(${}_{\textbf{6}}[{}^1S_0]$)
represents $(cc)$-diquark is in S-wave ${ }^{3} S_{1}$(${ }^{1} S_0$) and in the $\overline{\textbf{3}}$($\textbf{6}$) color state, while $h_{\bar{\textbf{3}}}$($h_{\textbf{6}}$) depicts its nonperturbative transiton probability into the baryon.
Extensive theoretical investigations have delved into the production of doubly heavy baryons \cite{Baranov:1995rc, Berezhnoy:1996an, Jiang:2012jt, Jiang:2013ej, Chen:2014frw, Yang:2014ita, Yang:2014tca, Martynenko:2013eoa, Zheng:2015ixa, Bi:2017nzv, Sun:2020mvl, Chen:2014hqa, Chen:2019ykv, Chen:2018koh, Martynenko:2014ola, Koshkarev:2016acq, Koshkarev:2016rci, Groote:2017szb, Berezhnoy:2018bde, Brodsky:2017ntu, Berezhnoy:2018krl, Wu:2019gta, Qin:2020zlg, Niu:2018ycb, Niu:2019xuq, Zhang:2022jst, Luo:2022jxq, Luo:2022lcj, Ma:2022cgt}. These investigations encompass direct production in $pp$, $ep$, $\gamma\gamma$ and $e^+e^-$ collisions, as well as indirect production via the decays of Higgs bosons, $W$ bosons, $Z$ bosons, and top quarks. A dedicated generator, GENXICC \cite{Chang:2007pp, Chang:2009va, Wang:2012vj}, has been developed to simulate hadroproduction in $pp$ collisions.

The $e^+e^-$ collider offers two primary avenues for the direct production of the doubly heavy baryon $\Xi_{QQ^{'}}$: production through $e^+e^-$ annihilation and via the photoproduction mechanism. In this work, $\Xi_{QQ^{'}}$ represents the baryon $\Xi_{QQ^{'}q}$, where $Q(Q^{'})$ stands for either a charm ($c$) or bottom ($b$) quark, and $q$ corresponds to an up ($u$), down ($d$), or strange ($s$) quark.
As for the photoproduction, $\Xi_{QQ^{'}}$ can be produced via direct photon-photon fusion such as $\gamma+\gamma \rightarrow \Xi_{QQ^{'}} +\bar{Q}+\bar{Q^{'}}$.
The collision photon may originate from either the bremsstrahlung of the initial $e^+e^-$ particles or from the process of laser back-scattering with $e^+e^-$.
In addition to direct photoproduction, there are also processes called resolved photoproduction \cite{Klasen:2001cu}, where the photon undergoes resolution, and its parton participates in the ensuing hard processes.
These resolved photoproduction channels share the same order of perturbative expansion as the direct approach, necessitating their inclusion in calculations. Noteworthy earlier studies \cite{Klasen:2001cu, Li:2009zzu, Zhan:2020ugq, Zhan:2021dlu, Zhan:2022nck, Zhan:2022etq} have indicated that these resolved channels tend to dominate the photoproduction of heavy quarkonium at $e^+e^-$ colliders.
Several next-generation $e^+e^-$ colliders have been proposed, including the FCC-ee \cite{FCC:2018evy}, the CEPC \cite{CEPCStudyGroup:2018rmc, CEPCStudyGroup:2018ghi}, and the ILC \cite{ILC:2007bjz, Erler:2000jg}. Designed to operate at varying high collision energies, along with unprecedented luminosities, these potent $e^+e^-$ colliders hold the potential to serve as exceptional platforms for diverse research subjects.

This study is anchored in the NRQCD framework, where we investigate the photoproduction of $\Xi_{QQ^{'}}$ at future $e^+e^-$ colliders. In addition to the direct photoproduction channel $\gamma + \gamma \rightarrow \Xi_{QQ^{'}} +\bar{Q}+\bar{Q^{'}}$, we also incorporate the resolved photoproduction processes, encompassing $\gamma + g \rightarrow \Xi_{QQ^{'}} +\bar{Q}+\bar{Q^{'}}$, $g + g \rightarrow \Xi_{QQ^{'}} +\bar{Q}+\bar{Q^{'}}$, and $q + \bar{q} \rightarrow \Xi_{QQ^{'}} +\bar{Q}+\bar{Q^{'}}$, where $q=u,d,s$. Section~\ref{sec:2} elucidates the calculation's formulation, while Section~\ref{sec:3} presents numerical outcomes and ensuing discussions. Section~\ref{sec:4} provides a succinct summary.

~\\

\begin{figure}
	\includegraphics[width=.4\textwidth]{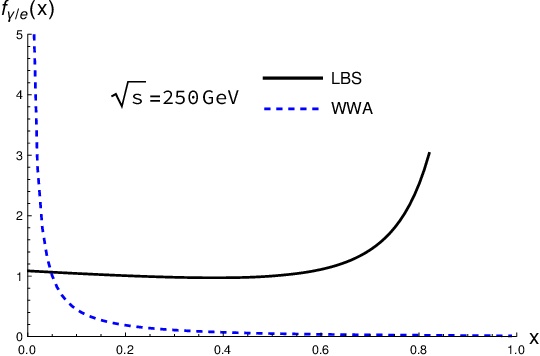}
	\caption{\label{fig:fre} The energy spectra of the LBS photon and the WWA photon.}
\end{figure}
\begin{figure*}[t]
	\centering
	\includegraphics[width=.8\textwidth]{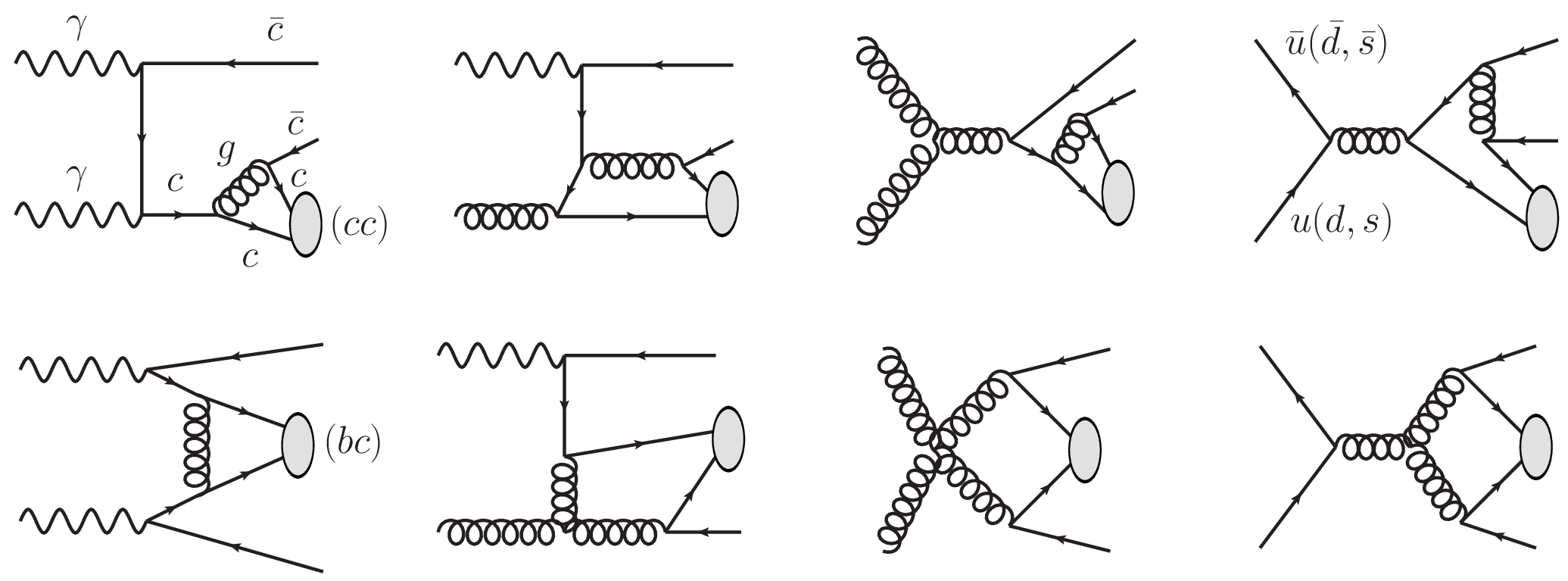}
	\caption{\label{fig:diag} Some typical Feynman diagrams for calculating the partonic cross section $\hat{\sigma}$ of $\Xi_{QQ^{'}}$ photoproduction. The diagrams are drawn by JaxoDraw~\cite{Binosi:2003yf}.}
\end{figure*}

\section{\label{sec:2}Formulation}

Based on the NRQCD factorization framework, the photoproduction cross section of $\Xi_{QQ^{'}}$ at the $e^+e^-$ collider can be expressed as,
\begin{eqnarray}
	&&\mathrm{d} \sigma\left(e^{+} e^{-} \rightarrow e^{+} e^{-} \Xi_{QQ^{'}}+\bar{Q}+\bar{Q^{'}}\right)\nonumber\\
	&&= \int \mathrm{d} x_{1} f_{\gamma / e}\left(x_{1}\right) \int \mathrm{d} x_{2} f_{\gamma / e}\left(x_{2}\right)\nonumber\\
	&& \times\sum_{i, j} \int \mathrm{d} x_{i} f_{i / \gamma}\left(x_{i}\right) \int \mathrm{d} x_{j} f_{j / \gamma}\left(x_{j}\right) \nonumber\\
	&&\times  \sum_{n} \mathrm{~d} \hat{\sigma}(i j \rightarrow (QQ^{'})[n]+\bar{Q}+\bar{Q^{'}})\left\langle \mathcal{O}^{\Xi_{QQ^{'}}}[n]\right\rangle.
\end{eqnarray}
Here $f_{\gamma/e}(x)$ is the energy spectrum of the photon.
$f_{i/\gamma}$($i=\gamma,g,u,d,s$) represents the Gl\"uck-Reya-Schienbein (GRS) distribution function of parton $i$ in photon~\cite{Gluck:1999ub}.
$f_{\gamma/\gamma}(x)=\delta(1-x)$ is for the direct photoproduction process.
$\mathrm{~d} \hat{\sigma}(i j \rightarrow (QQ^{'})[n]+\bar{Q}+\bar{Q^{'}})$ is the differential partonic cross section, which is calculated perturbatively.
For baryon $\Xi_{cc}$ and $\Xi_{bb}$, $n={}_{\bar{\textbf{3}}}[{}^3S_1]$ and ${}_{\textbf{6}}[{}^1S_0]$.
For $\Xi_{bc}$, $n={}_{\bar{\textbf{3}}}[{}^3S_1]$, ${}_{\textbf{6}}[{}^1S_0]$, ${}_{\bar{\textbf{3}}}[{}^1S_0]$ and ${}_{\bar{\textbf{3}}}[{}^1S_0]$.
$\langle{\cal O}^{\Xi_{QQ^{'}}}[n]\rangle=h_n$ is the long distance matrix element(LDME).
People usually employ potential model, mimicking the heavy quarkonium case, introduce and relate a wave function to $h_{\bar{\textbf{3}}}$ \cite{Falk:1993gb,Kiselev:1994pu,Bagan:1994dy,Baranov:1995rc,Berezhnoy:1998aa},
\begin{equation}
	h_{\bar{\textbf{3}}}\simeq|\Psi_{QQ^{'}}(0)|^2.
	\label{eq:h3}
\end{equation}
As for $h_{\textbf{6}}$, there is no such relation and it is set to equal to $h_{\bar{\textbf{3}}}$ for simplicity.
This assumption is grounded in NRQCD's power counting with respect to $v_c$, where both $h_{\textbf{6}}$ and $h_{\bar{\textbf{3}}}$ hold equivalent orders \cite{Ma:2003zk}. 
According to NRQCD, the bound state $\Xi_{Q c}$ can be expanded into a series of Fock states,
\begin{equation}
	\label{eq:fock}
	\begin{aligned}
		\left|\Xi_{Q Q}\right\rangle= & c_1(v)|(Q Q) q\rangle+c_2(v)|(Q Q) q g\rangle \\
		& +c_3(v)|(Q Q) q g g\rangle+\cdots.
	\end{aligned}
\end{equation}
Since a light quark can readily emit gluons, the constituents in Eq.~(\ref{eq:fock}) hold equivalent importance, specifically, $c_1\sim c_2 \sim c_3$. 
Consider a $QQ$ pair in the ${}_{\bar{\textbf{3}}}[{}^3S_1]$ state; one of the heavy quarks can emit a gluon without altering the spin of the heavy quark. Subsequently, this gluon undergoes a splitting into a pair of light quarks $q\bar{q}$, permitting the heavy $QQ$ pair to engage with the light $q$ to compose $\Xi_{Q Q}$. Similarly, for a $QQ$ pair in the ${}_{\textbf{6}}[{}^1S_0]$ state, one of the heavy quarks can emit a gluon that retains the spin of the heavy quark unchanged. This emitted gluon then segregates into a light $q\bar{q}$ pair, and the light quarks also exhibit a propensity for gluon emission. Consequently, this heavy $QQ$ pair can capture a light quark and a gluon to assemble into $\Xi_{Q Q}$.
This elucidates why $h_{\textbf{6}}$ and $h_{\bar{\textbf{3}}}$ hold the same order in $v_c$. For simplicity, we assume $h_{\textbf{6}}=h_{\bar{\textbf{3}}}$ in the ensuing computations. Notably, the LDMEs serve as overarching parameters beyond the perturbative components, implying that the outcomes can be readily refined upon acquisition of novel LDMEs.

As previously mentioned, the $e^+e^-$ collider presents two primary sources of initial photons. The first emanates from the bremsstrahlung of the initial $e^+e^-$ pairs, and its energy distribution can be well delineated within the Weiz\"acker-Williams approximation (WWA)~\cite{Frixione:1993yw},
\begin{eqnarray}
	f_{\gamma/e}(x) &=& \frac{\alpha}{2\pi}\Bigg[\frac{1 + (1 - x)^2}{x} {\rm log}\frac{Q^2_{\rm max}}{Q^2_{\rm min}} \nonumber\\
	&&+2m_e^2x\left(\frac{1}{Q^2_{\rm max}}
	-\frac{1}{Q^2_{\rm min}}\right)\Bigg],
\end{eqnarray}
where $x=E_{\gamma} / E_{e}$ represents the fraction of longitudinal momentum carried by the photon, while $\alpha$ denotes the electromagnetic fine structure constant.
$Q^2_{\rm min} = m_e^2 x^2/(1-x)$ and $Q^2_{\rm max} = (E\theta_c)^2(1-x) + Q^2_{\rm min}$, with $\theta_c=32\mathrm{~mrad}$ defining the maximum scattered angular cut to ensure photon to be real. Here, $E=E_e=\sqrt{s}/2$ reflects the collision energy, defined as $\sqrt{s}$.

Another source is from the laser back-scattering (LBS) with $e^+e^-$ and its spectrum function is~\cite{Ginzburg:1981vm},
\begin{eqnarray}
	f_{\gamma/e}(x)=\frac{1}{N}\left[1-x+\frac{1}{1-x}-4 r(1-r)\right],
\end{eqnarray}
where $r=x /\left[x_{m}(1-x)\right]$, and the normalization factor,
\begin{eqnarray}
	N&=&\left(1-\frac{4}{x_{m}}-\frac{8}{x_{m}^{2}}\right) \log(1+x_m)\nonumber\\
	&&+\frac{1}{2}+\frac{8}{x_{m}}-\frac{1}{2 (1+x_m)^{2}}.
\end{eqnarray}
Here $x_{m}=4 E_{e} E_{l} \cos ^{2} \frac{\theta}{2}$, $E_e$ and $E_l$ are the energies of incident electron and laser beams, respectively. $\theta$ is the angle between them. The energy of the LBS photon is restricted by
\begin{eqnarray}
	0 \leq x \leq \frac{x_{m}}{1+x_{m}},
\end{eqnarray}
with optimal value of $x_m$ being $4.83$~\cite{Telnov:1989sd}. These two spectra have quite different behaviors as shown in Fig.~\ref{fig:fre}.

Some typical Feynman diagrams for calculating the partonic cross sections are shown in Fig.~\ref{fig:diag}.
The well-established system Feynman Diagram Calculation (FDC)~\cite{Wang:2004du} is used in the analytical and numerical calculations, where the standard projection method~\cite{Bodwin:2002cfe} is employed to deal with the amplitudes.

\begin{table}[b]
	\caption{\label{tab:cross-section-1}The integrated cross sections (in unit of fb) under default inputs for the photoproduction of $\Xi_{QQ^{'}}$ via the LBS photon and the WWA photon (in brackets), respectively. Three typical collision energies are taken as example and intermediate diquark at various spin- and color-configurations are listed.}
	\begin{ruledtabular}
		\begin{tabular}{cccc}
			$\sqrt{S}(\mathrm{GeV})$ & 250 & 500 & 1000\\
			\colrule
			$(cc)_{\bar{\textbf{3}}}[{}^3S_1]$ & 733.68(62.75) & 801.90(111.74) & 1100.09(182.73) \\
			$(cc)_{\textbf{6}}[{}^1S_0]$ & 65.65(2.79) & 75.44(5.35) & 105.89(9.35)\\
			$(bc)_{\bar{\textbf{3}}}[{}^3S_1]$ & 26.27(0.85) & 30.44(1.73) & 44.85(3.14)\\
			$(bc)_{\textbf{6}}[{}^1S_0]$ & 5.72(0.18) & 6.46(0.36) & 9.44(0.66)\\
			$(bc)_{\textbf{6}}[{}^3S_1]$ & 13.14(0.43) & 15.22(0.86) & 22.43(1.57)\\	
			$(bc)_{\bar{\textbf{3}}}[{}^1S_0]$ & 11.45(0.35) & 12.91(0.72) & 18.88(1.32)\\
			$(bb)_{\bar{\textbf{3}}}[{}^3S_1]$ & 1.25(0.02) & 1.56(0.05) & 2.44(0.10) \\
			$(bb)_{\textbf{6}}[{}^1S_0]$ & 0.09(0.0008) & 0.14(0.002) & 0.23(0.005)\\
		\end{tabular}
	\end{ruledtabular}
\end{table}

\section{\label{sec:3}Numerical results and discussions}

In the calculation, we take the wave functions at the origin as\cite{Bagan:1994dy} $|\Psi_{cc}(0)|^2=0.039\mathrm{~GeV^3}$, $|\Psi_{bb}(0)|^2=0.152\mathrm{~GeV^3}$ and $|\Psi_{bc}(0)|^2=0.065\mathrm{~GeV^3}$. 
For consistency, we also fix the quark masses as they are given in Ref.\cite{Bagan:1994dy}: $m_c=M_{\Xi_{cc}}/2=1.8\mathrm{~GeV}$ and $m_b=M_{\Xi_{bb}}/2=5.1\mathrm{~GeV}$.
The fine structure constant is set to be $\alpha =1/137$.
As for the strong coupling constant, the one-loop running formulation is employed. The renormalization scale is by default taken as the transverse mass of $\Xi_{QQ^{'}}$, $\mu=\sqrt{M^2_{\Xi_{QQ^{'}}}+p^2_t}$ with $p_t$ representing its transverse momentum.

\begin{table*}
	\caption{\label{tab:cross-section-2}
		The integrated cross sections (in unit of fb) of different channels of the photoproduction of $\Xi_{QQ^{'}}$ via the LBS photon. Three typical collision energies, $250(500,1000)\mathrm{~GeV}$, are taken and contributions of all intermediate diquark states have been summed up.}
	\begin{ruledtabular}
		\begin{tabular}{ccccc}
			$channels$ & $\gamma+\gamma$ & $\gamma+g$ & $g+g$ & $q+\bar{q}$\\
			\colrule
			$\Xi_{cc}$ & $392.72 (173.84,70.13)$ & $402.37 (693.28,1112.50)$ & $3.22 (9.35,22.50)$ & $1.02(0.88,0.85)$ \\
			$\Xi_{bc}$ & $32.85 (16.02,6.96)$ & $23.00 (47.31,84.34)$ & $0.36 (1.37,3.99)$ & $0.37(0.33,0.30)$ \\
			$\Xi_{bb}$ & $0.80 (0.44,0.21)$ & $0.51 (1.20,2.33)$ & $0.0066 (0.033,0.11)$ & $0.019(0.018,0.017)$ \\
		\end{tabular}
	\end{ruledtabular}
\end{table*}
\begin{table*}
	\caption{\label{tab:cross-section-3}
		The integrated cross sections (in unit of fb) of different channels of the photoproduction of $\Xi_{QQ^{'}}$ via the WWA photon. Three typical collision energies, $250(500,1000)\mathrm{~GeV}$, are taken and  contributions of all intermediate diquark states have been summed up.}
	\begin{ruledtabular}
		\begin{tabular}{ccccc}
			$channels$ & $\gamma+\gamma$ & $\gamma+g$ & $g+g$ & $q+\bar{q}$\\
			\colrule
			$\Xi_{cc}$ & $62.88 (109.17,172.06)$ & $2.63 (7.83,19.77)$ & $0.0088 (0.042,0.16)$ & $0.023(0.047,0.085)$ \\
			$\Xi_{bc}$ & $1.71 (3.32,5.64)$ & $0.09 (0.34,1.03)$ & $0.00070 (0.0045,0.021)$ & $0.003(0.008,0.017)$ \\
			$\Xi_{bb}$ & $0.02 (0.04,0.08)$ & $0.0015 (0.0068,0.02)$ & $9.6\times10^{-6} (8.4\times10^{-5},4.8\times10^{-4})$ & $9.7\times10^{-5} (2.8\times10^{-4},6.2\times10^{-4})$ \\
		\end{tabular}
	\end{ruledtabular}
\end{table*}
\begin{figure*}
	\includegraphics[width=.24\textwidth]{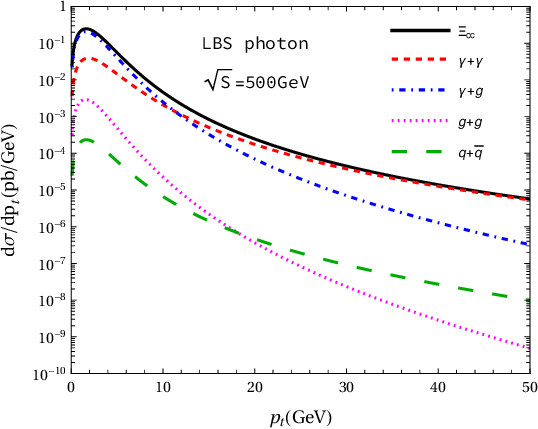}
	\includegraphics[width=.24\textwidth]{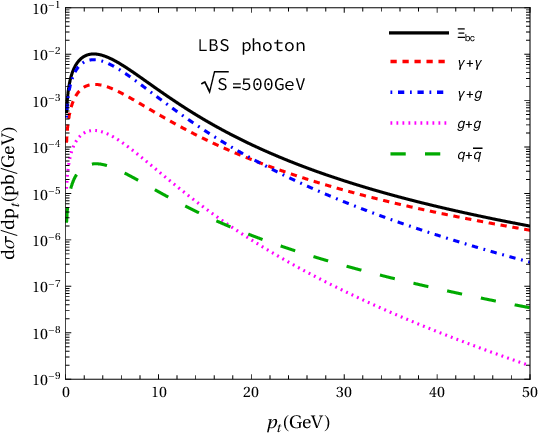}
	\includegraphics[width=.24\textwidth]{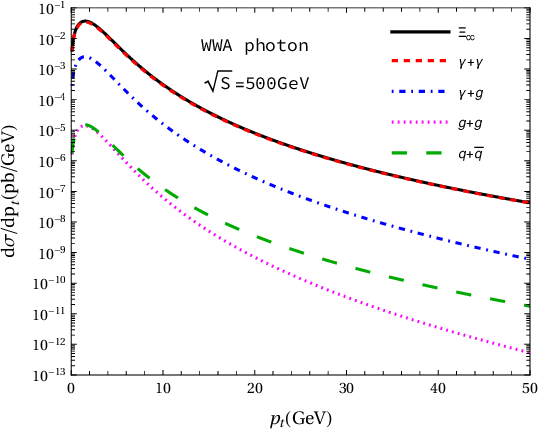}
	\includegraphics[width=.24\textwidth]{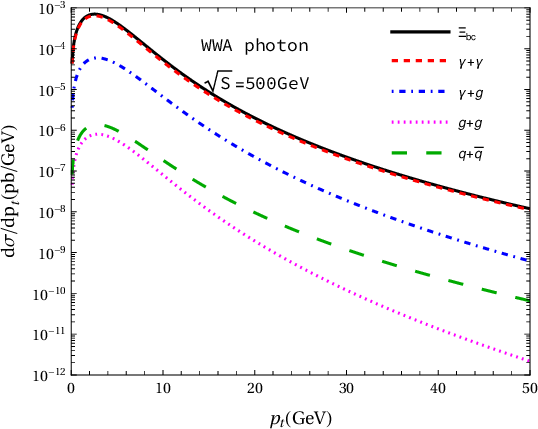}\\
			\includegraphics[width=.24\textwidth]{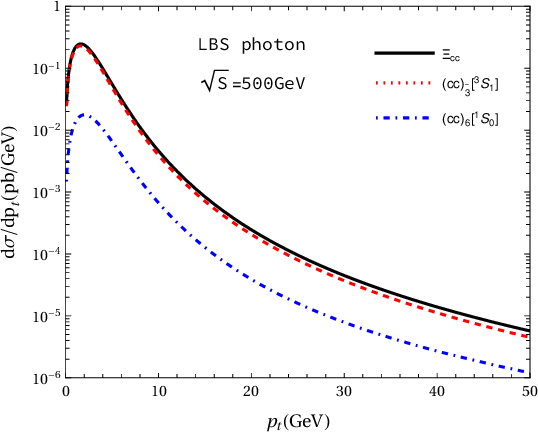}
	\includegraphics[width=.24\textwidth]{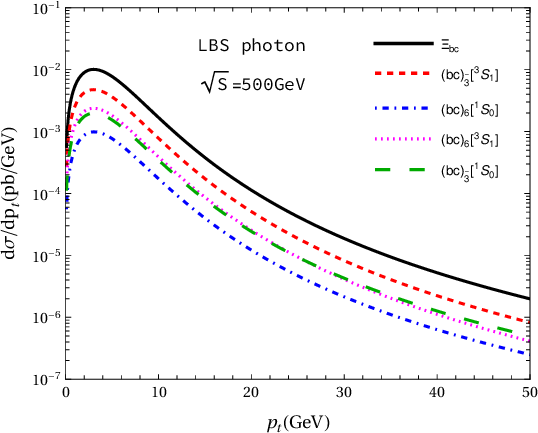}
	\includegraphics[width=.24\textwidth]{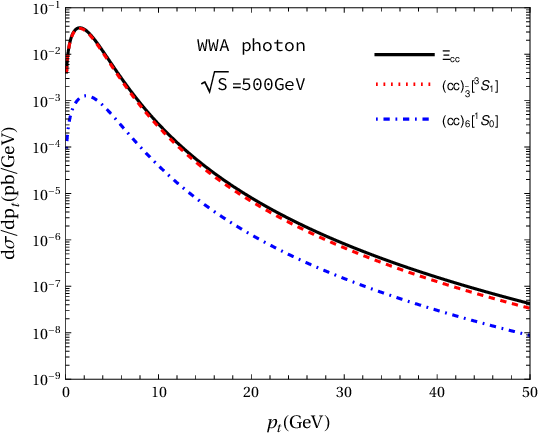}
	\includegraphics[width=.24\textwidth]{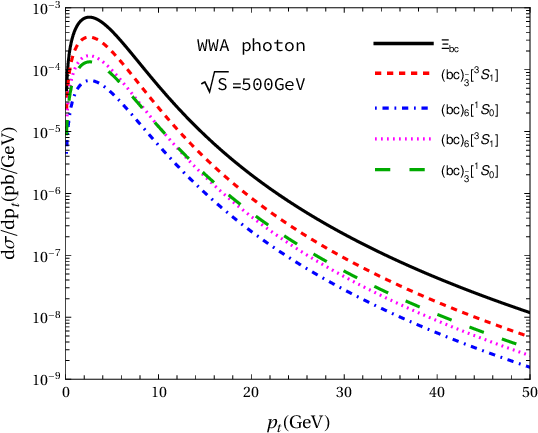}
	\caption{\label{fig:pt} 
	The $p_t$ distributions for $\Xi_{cc}$ and $\Xi_{b c}$ photoproduction under $\sqrt{S}=500\mathrm{~GeV}$ and default values of the parameters. Figures in the first row are for different channels and those in the second row are for different intermediate diquark states. The topmost curve in every figure is their summation. Two columns of figures on the left are for LBS photon and those on the right are for WWA photon.}
\end{figure*}

\begin{figure*}
	\includegraphics[width=.24\textwidth]{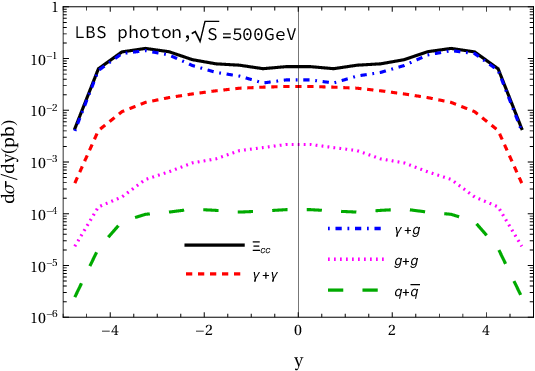}
	\includegraphics[width=.24\textwidth]{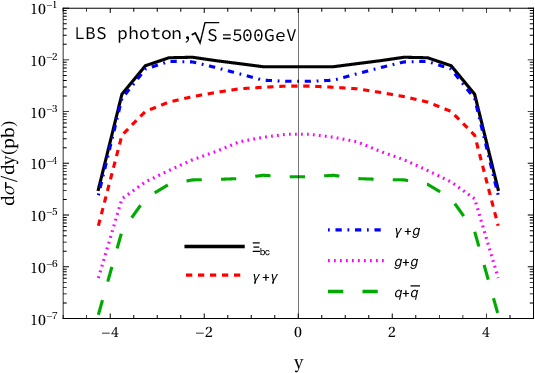}
	\includegraphics[width=.24\textwidth]{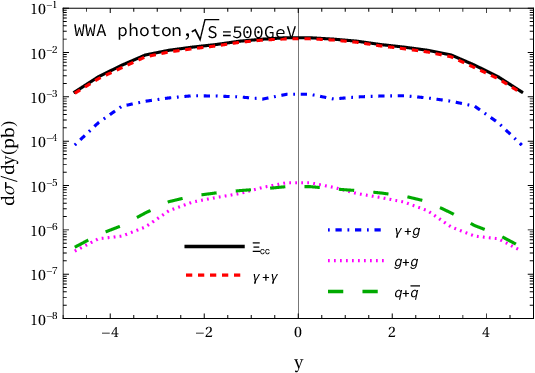}
	\includegraphics[width=.24\textwidth]{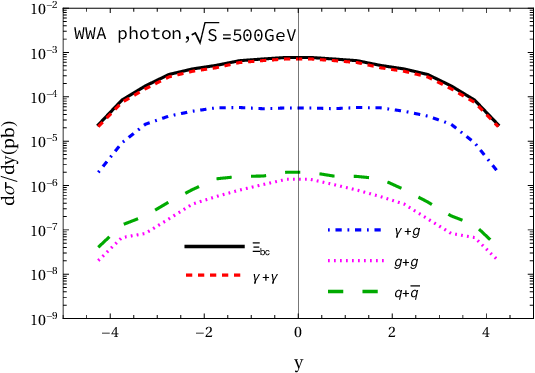}\\
	\includegraphics[width=.24\textwidth]{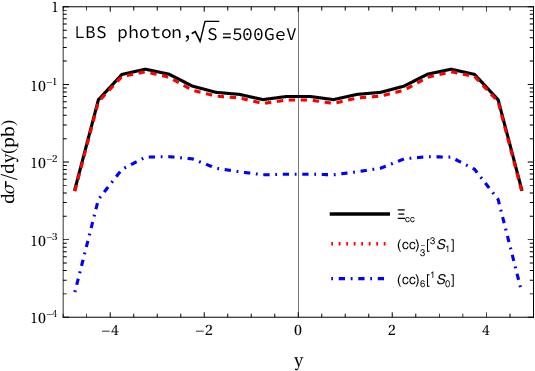}
	\includegraphics[width=.24\textwidth]{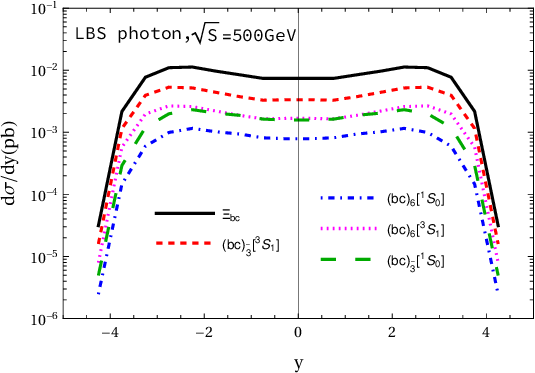}
	\includegraphics[width=.24\textwidth]{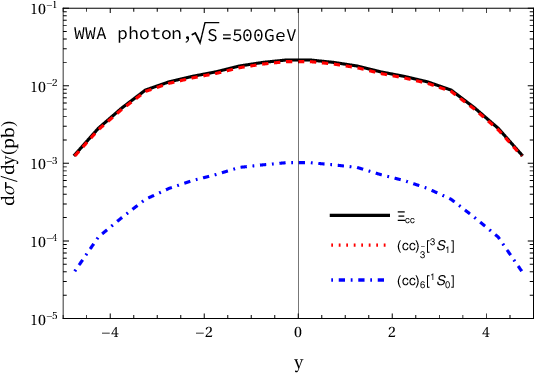}
	\includegraphics[width=.24\textwidth]{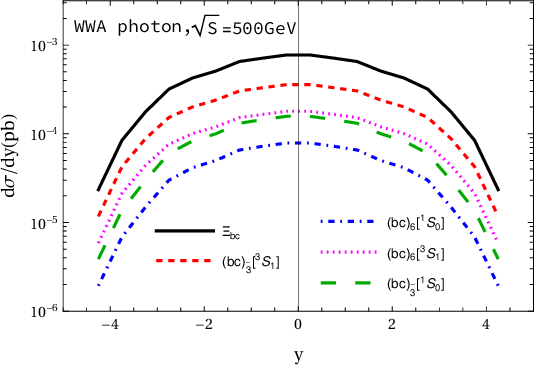}
	\caption{\label{fig:y}
	The $y$ distributions for $\Xi_{cc}$ and $\Xi_{b c}$ photoproduction under $\sqrt{S}=500\mathrm{~GeV}$ and default values of the parameters. Figures in the first row are for different channels and those in the second row are for different intermediate diquark states. The topmost curve in every figure is their summation. Two columns of figures on the left are for LBS photon and those on the right are for WWA photon.}
\end{figure*}

For collision energies of $\sqrt{S}=250$, $500$, and $1000\mathrm{GeV}$, along with the default inputs, Table~\ref{tab:cross-section-1} illustrates the cross section for the photoproduction of $\Xi_{QQ^{'}}$ using both LBS and WWA photons (in brackets). This table includes a range of spin and color configurations. The results highlight a consistent trend: as collision energy increases, all cross sections exhibit growth,
for the LBS photoproduction,
\begin{equation}
	\begin{aligned}
		& \left.\sigma_{\Xi_{c c}}\right|_{250 \mathrm{GeV}}:\left.\sigma_{\Xi_{c c}}\right|_{500 \mathrm{GeV}}:\left.\sigma_{\Xi_{c c}}\right|_{1 \mathrm{TeV}} \simeq 1: 1.10: 1.60, \\
		& \left.\sigma_{\Xi_{b c}}\right|_{250 \mathrm{GeV}}:\left.\sigma_{\Xi_{b c}}\right|_{500 \mathrm{GeV}}:\left.\sigma_{\Xi_{b c}}\right|_{1 \mathrm{TeV}} \simeq 1:1.15:1.69, \\
		& \left.\sigma_{\Xi_{b b}}\right|_{250 \mathrm{GeV}}:\left.\sigma_{\Xi_{b b}}\right|_{500 \mathrm{GeV}}:\left.\sigma_{\Xi_{b b}}\right|_{1 \mathrm{TeV}} \simeq 1: 1.27: 1.99 ,
	\end{aligned}
\end{equation}
and for the WWA photoproduction,
\begin{equation}
	\begin{aligned}
		& \left.\sigma_{\Xi_{c c}}\right|_{250 \mathrm{GeV}}:\left.\sigma_{\Xi_{c c}}\right|_{500 \mathrm{GeV}}:\left.\sigma_{\Xi_{c c}}\right|_{1 \mathrm{TeV}} \simeq 1: 1.79: 2.93, \\
		& \left.\sigma_{\Xi_{b c}}\right|_{250 \mathrm{GeV}}:\left.\sigma_{\Xi_{b c}}\right|_{500 \mathrm{GeV}}:\left.\sigma_{\Xi_{b c}}\right|_{1 \mathrm{TeV}} \simeq 1: 2.03: 3.70, \\
		& \left.\sigma_{\Xi_{b b}}\right|_{250 \mathrm{GeV}}:\left.\sigma_{\Xi_{b b}}\right|_{500 \mathrm{GeV}}:\left.\sigma_{\Xi_{b b}}\right|_{1 \mathrm{TeV}} \simeq 1: 2.50: 5.05 .
	\end{aligned}
\end{equation}
The spin- and color-configurations also give different contributions to the total cross section and under $\sqrt{S}=500\mathrm{~GeV}$ for the LBS photoproduction,
\begin{equation}
	\begin{aligned}
		& \sigma_{(c c)_{\overline{3}}\left[{ }^3 S_1\right]}: \sigma_{(c c)_{\mathbf{6}}\left[{ }^1 S_0\right]} \simeq 10.63: 1, \\
		& \sigma_{\left.(b c)_{\overline{3}}{ }^3 S_1\right]}: \sigma_{(b c)_{\mathbf{6}}\left[{ }^1 S_0\right]}: \sigma_{(b c)_{\mathbf{6}}\left[{ }^3 S_1\right]}: \sigma_{(b c)_{\overline{3}}\left[{ }^3 S_1\right]} \\
		& \simeq 4.71: 1: 2.36: 2.00, \\
		& \sigma_{(b b)_{\overline{3}}\left[{ }^3 S_1\right]}: \sigma_{(b b)_{\mathbf{6}}\left[{ }^1 S_0\right]} \simeq 11.14: 1.
	\end{aligned}
\end{equation}
The cross sections via the LBS photoproduction are much larger than those of the WWA. This is due to the quite different spectra functions of the photon as shown in Fig.~\ref{fig:fre}.
When assuming an integrated luminosity of $\mathcal{O}(10^4)\mathrm{~fb^{-1}}$ at future $e^+e^-$ colliders and aggregating contributions from all diquark configurations, approximately $8.8\times10^6$ ($1.2\times10^6$) $\Xi_{cc}$, $6.5\times10^5$ ($3.7\times10^4$) $\Xi_{bc}$, and $1.7\times10^4$ ($4.9\times10^2$) $\Xi_{bb}$ would be generated via LBS (WWA) photons, given a collision energy of $\sqrt{S}=500\mathrm{~GeV}$.
The actual number of experimentally reconstructed events is significantly affected both by the decay rate of $\Xi_{QQ^{'}}$ and the experimental reconstruction efficiency.
Considering, for instance, the cascade decay $\Xi_{c c}^{++} \rightarrow \Lambda_c^{+} K^{-} \pi^{+} \pi^{+} \simeq 10\%$\cite{Yu:2017zst} and $\Lambda_c^{+} \rightarrow p K^{+} \pi^{+} \simeq 5\%$\cite{Aaij:2013voa}, and accounting for the experiment's reconstruction efficiency, the event count is expected to diminish by approximately three orders of magnitude.
Consequently the photoproduction at future $e^+e^-$ colliders could provides good opportunity to study $\Xi_{cc}$, while there would be not enough events for $\Xi_{bb}$.

Table~\ref{tab:cross-section-2} lists the contributions from different channels for the LBS photon. With the increase of collision energy, the cross sections of $\gamma+\gamma$ and $q+\bar{q}$ channels decrease, while those of the other two become larger. The $\gamma+g$ channels provide very important contributions at all these three collision energy and even dominant when the collision energy goes larger.
The double resolved channels, $g+g$ and $q+\bar{q}$, give very tiny productions which can be ignored compared with other two.
Consequently for the LBS photon, the resolved photoproduction channel($\gamma+g$) at future $e^+e^-$ colliders should be taken into account in the theoretical investigation.
Conversely, in WWA photoproduction, the relative importance among various channels markedly deviates from that in LBS, as highlighted in Table~\ref{tab:cross-section-3}. The cross sections across all channels burgeon with increasing collision energy. While the direct $\gamma+\gamma$ channels consistently dominate, the contributions from other channels remain modest, and in certain cases, negligible.

Fig.~\ref{fig:pt} illustrates the transverse momentum distributions of $\Xi_{cc}$ and $\Xi_{bc}$ photoproduction, with separate representations of contributions from different channels and intermediate diquark states. Each $p_t$ distribution exhibits a discernible peak around several $\mathrm{GeV}$, transitioning into a logarithmic decrease in the high $p_t$ range.
Throughout the entire $p_t$ spectrum, the ${}_{\bar{\textbf{3}}}[{}^3S_1]$ configurations consistently hold prominence, rendering contributions from ${}_{\textbf{6}}[{}^1S_0]$ states inconsequential. Specifically, in LBS photoproduction, the $\gamma+g$ channels dominate the lower $p_t$ region, yielding the baton to the $\gamma+\gamma$ channels as the $p_t$ value increases significantly.
However, in practical experiments, there maybe not enough events in large $p_t$ region to make precise measurements and consequently the single resolved channel $\gamma+g$ must be considered in the calculation of photoproduction. For the WWA case, the direct channel $\gamma+\gamma$ is always predominant in whole $p_t$ region.

Fig.~\ref{fig:y} portrays the rapidity ($y$) distributions of $\Xi_{cc}$ and $\Xi_{bc}$ photoproduction. In the LBS scenario, the curves exhibit distinctive patterns across the central rapidity region, attributed to the prevalence of the $\gamma+g$ channel. In contrast to the transverse momentum ($p_t$) distribution, the $\gamma+g$ and $\gamma+\gamma$ channels do not intersect throughout the entire $y$ range for a collision energy of $\sqrt{S}=500\mathrm{~GeV}$.
Comparatively, the rapidity distributions of the WWA photoproduction appear conventional when juxtaposed with those of the LBS scenario.

\begin{table}
	\caption{\label{tab:uncer-mc}The total cross sections (in unit of fb) under different $m_c$ for the $\Xi_{QQ^{'}}$ production via the LBS photon and WWA photon (in brackets) at $\sqrt{S}=500\mathrm{~GeV}$, respectively. All sub-channels have been summed up.}
	\begin{ruledtabular}
	\begin{tabular}{cccc}
		$m_c(\mathrm{GeV})$ & 1.7 & 1.8 & 1.9\\
		\colrule
		$(cc)_{\bar{\textbf{3}}}[{}^3S_1]$ & 1104.64(159.47) & 801.90(111.74) & 594.56(79.88) \\
		$(cc)_{\textbf{6}}[{}^1S_0]$ & 104.07(7.66) & 75.44(5.35) & 55.80(3.81)\\
		$(bc)_{\bar{\textbf{3}}}[{}^3S_1]$ & 36.10(2.05) & 30.44(1.73) & 25.86(1.47)\\
		$(bc)_{\textbf{6}}[{}^1S_0]$ & 7.74(0.43) & 6.46(0.36) & 5.45(0.31)\\
		$(bc)_{\textbf{6}}[{}^3S_1]$ & 18.05(1.03) & 15.22(0.86) & 12.93(0.73)\\	
		$(bc)_{\bar{\textbf{3}}}[{}^1S_0]$ & 15.47(0.85) & 12.91(0.72) & 10.89(0.62)\\
	\end{tabular}
\end{ruledtabular}
\end{table}
\begin{table}
	\caption{\label{tab:uncer-mb}The total cross sections (in unit of fb) under different $m_b$ for the $\Xi_{QQ^{'}}$ production via the LBS photon and WWA photon (in brackets) at $\sqrt{S}=500\mathrm{~GeV}$, respectively. All sub-channels have been summed up.}
	\begin{ruledtabular}
	\begin{tabular}{cccc}
	$m_b(\mathrm{GeV})$ & 4.9 & 5.1 & 5.3 \\
	\colrule
	$(bc)_{\bar{\textbf{3}}}[{}^3S_1]$ & 35.25(2.05) & 30.44(1.73) & 26.39(1.47)\\
	$(bc)_{\textbf{6}}[{}^1S_0]$ & 7.42(0.43) & 6.46(0.36) & 5.62(0.31)\\
	$(bc)_{\textbf{6}}[{}^3S_1]$ & 17.63(1.02) & 15.22(0.86) & 13.19(0.73)\\	
	$(bc)_{\bar{\textbf{3}}}[{}^1S_0]$ & 14.84(0.86) & 12.91(0.72) & 11.23(0.61)\\
\end{tabular}
	\end{ruledtabular}
\end{table}
\begin{table}[!t]
	\caption{\label{tab:uncer-mu}The total cross sections (in unit of fb) under various $\mu(={\cal C}\sqrt{M^2_{\Xi_{QQ^{'}}}+p^2_t}$ with ${\cal C}=0.5,1,2$) for the photoproduction of $\Xi_{QQ^{'}}$ via the LBS photon and WWA photon (in brackets) at $\sqrt{S}=500\mathrm{~GeV}$, respectively. All sub-channels have been summed up.}
	\begin{ruledtabular}
		\begin{tabular}{cccc}
			${\cal C}$ & $0.5$ & $1.0$ & $2.0$\\
			\colrule
		$(cc)_{\bar{\textbf{3}}}[{}^3S_1]$ & 960.64(184.36) & 801.90(111.74) & 684.06(76.93) \\
		$(cc)_{\textbf{6}}[{}^1S_0]$ & 91.82(8.53) & 75.44(5.35) & 64.48(3.77)\\
		$(bc)_{\bar{\textbf{3}}}[{}^3S_1]$ & 37.56(2.54) & 30.44(1.73) & 25.07(1.26)\\
		$(bc)_{\textbf{6}}[{}^1S_0]$ & 8.02(0.53) & 6.46(0.36) & 5.32(0.26)\\
		$(bc)_{\textbf{6}}[{}^3S_1]$ & 18.78(1.27) & 15.22(0.86) & 12.53(0.64)\\	
		$(bc)_{\bar{\textbf{3}}}[{}^1S_0]$ & 16.03(1.06) & 12.91(0.72) & 10.63(0.53)\\
\end{tabular}
	\end{ruledtabular}
\end{table}

Finally, we engage in a concise discussion on the theoretical uncertainties within our calculations, stemming from three primary sources: the heavy quark masses, the renormalization scale $\mu$, and the LDMEs. Notably, uncertainties arising from $h_{\bar{\textbf{3}}}$ and $h_{\textbf{6}}$ are omitted due to the absence of reported errors in the literature. As previously indicated, these coefficients represent overall factors, and their impact on production outcomes can be readily refined with more accurate values.
Table~\ref{tab:uncer-mc} illustrates the effects of varying $m_c=1.8\pm0.1\mathrm{~GeV}$ while holding $m_b=5.1\mathrm{GeV}$ and $\mu=\sqrt{M^2_{\Xi_{QQ^{'}}}+p^2_t}$ constant. Similarly, Table\ref{tab:uncer-mb} presents uncertainties resulting from $m_b=5.1\pm0.2\mathrm{~GeV}$ alongside $m_c=1.8\mathrm{GeV}$ and $\mu=\sqrt{M^2_{\Xi_{QQ^{'}}}+p^2_t}$. From these tables, even slight deviations in heavy quark mass can yield significant fluctuations in cross-section values. For instance, in Table~\ref{tab:uncer-mc}, the cross section for $(cc)_{\bar{\textbf{3}}}[{}^3S_1]$ varies by approximately 46\% for a mere 12\% alteration in $m_c$.
This pronounced sensitivity is illuminated when examining the pertinent Feynman diagrams, as exemplified in Fig.~\ref{fig:diag}.
For photoproduction of $\Xi_{QQ^{'}}$ considered here, the final particles in the short-distance processes encompass solely $Q$ and $\bar{Q}$ ($Q=c$ or $b$), with corresponding internal lines comprising exclusively $Q$ and gluon propagators. 
Hence, the profound influence of heavy quark masses on cross section appears reasonable. 
Let us take the second diagram($\gamma+g\rightarrow c+c+\bar{c}+\bar{c}$) in the second row in Fig.~\ref{fig:diag} as an example, which is one of the predominant diagrams, to illustrate the strong dependence on $m_c$.
The squared invariant mass of the gluon propagator that attached to the final $c\bar{c}$ pair is $k^2=(p_c+p_{\bar{c}})^2$.
Its dominant region in phase space integration is near the threshold, i.e., $k^2\sim4m_c^2$.
Consequently, when $m_c$ changes from $1.7\mathrm{~GeV}$ to $1.9\mathrm{~GeV}$, $1/(k^2)^2$ changes by about $36\%$.
Table~\ref{tab:uncer-mu} assesses the sensitivity to the renormalization scale ($\mu={\cal C}\sqrt{M^2_{\Xi_{QQ^{'}}}+p^2_t}$, with ${\cal C}=0.5,1,2$), considering fixed values for $m_c=1.8\mathrm{~GeV}$ and $m_b=5.1\mathrm{~GeV}$. Evidently, substantial dependence on the renormalization scale is evident, potentially signifying the relevance of next-to-leading order corrections in $\alpha_s$. As we confront real-world measurements in the future, high-order calculations become imperative.
Considering above uncertainties, the results in our leading order calculation may fluctuate by about one order of magnitude.
Within this range of variability, the photoproduction rates of doubly heavy baryons remain appreciable.

\section{\label{sec:4}Summary}

In this work, we have investigated the $\Xi_{QQ^{'}}$ photoproduction within the framework of non-relativistic QCD specifically focusing on future $e^+e^-$ colliders. 
The investigation encompasses two distinct sources of initial photons: the LBS photon and the WWA photon.
Two $(cc(bb))$-diquark configurations, ${}_{\bar{\textbf{3}}}[{}^3S_1]$ and ${}_{\textbf{6}}[{}^1S_0]$, and four $(bc)$-diquark configurations, $(bc)_{\bar{\textbf{3}}}[{}^3S_1]$, $(bc)_{\textbf{6}}[{}^1S_0]$, $(bc)_{\textbf{6}}[{}^3S_1]$ and $(bc)_{\bar{\textbf{3}}}[{}^1S_0]$, are considered.
Upon assuming $h_{\textbf{6}}=h_{\bar{\textbf{3}}}$, the results demonstrate ${}_{\bar{\textbf{3}}}[{}^3S_1]$ diquark state give dominant cross section,
while other intermediate states also provide notable contributions.
Importantly, beyond the direct photoproduction channel $\gamma+\gamma$, our study particularly integrates the resolved mechanisms, which are not fully considered in previous studies.
Numeric findings underscore the critical role of the single resolved photoproduction channel $\gamma+g$ in LBS photoproduction, while its significance diminishes in the WWA scenario.
If setting the integrated luminosity of future $e^+e^-$ collision as $\mathcal{O}(10^4)\mathrm{~fb^{-1}}$, there would be about $8.8\times10^6$ ($1.2\times10^6$) $\Xi_{cc}$ and $6.5\times10^5$ ($3.7\times10^4$) $\Xi_{bc}$ baryons to be generated via the LBS (WWA) photons respectively under the collision energy $\sqrt{S}=500\mathrm{~GeV}$.
While acknowledging the relatively considerable uncertainties inherent in our calculations, we anticipate that these findings could serve as a valuable preliminary exploration into photoproduction at prospective $e^+e^-$ colliders.

\begin{acknowledgments}

This work was supported in part by the Natural Science Foundation of China under Grants No. 12147116, No. 12175025, No. 12005028 and No. 12147102, by the China Postdoctoral Science Foundation under Grant No. 2021M693743 and by the graduate research and innovation foundation of Chongqing, China under Grant No.ydstd1912.

\end{acknowledgments}
%

\end{document}